\documentclass[runningheads]{llncs}
\usepackage{hyperref}
\usepackage{subcaption}
\usepackage{url}
\usepackage{graphicx}
\usepackage{siunitx}
\usepackage{amsmath}
\usepackage{amssymb}
\usepackage{tablefootnote}
\usepackage{listings}
\usepackage{xcolor}  
\usepackage{bsymb}
\usepackage[capitalise,noabbrev]{cleveref}
\usepackage{wrapfig}
\usepackage{enumitem}
\usepackage{algorithm}
\usepackage{algpseudocode}
\usepackage{mathtools}
\usepackage{todonotes}
\usepackage{framed}

\usepackage{enumitem}
\usepackage[misc]{ifsym}

\newlist{casesList}{enumerate}{2}
\setlist[casesList,1]{label=Case \arabic*:,ref=Case \arabic*}
\setlist[casesList,2]{label=(\alph*),ref=\thequestionsi(\alph*)}
\definecolor{orcidlogocol}{HTML}{A6CE39}
\usepackage{tikz}
\usetikzlibrary{svg.path}
\tikzset{
  orcidlogo/.pic={
    \fill[orcidlogocol] svg{M256,128c0,70.7-57.3,128-128,128C57.3,256,0,198.7,0,128C0,57.3,57.3,0,128,0C198.7,0,256,57.3,256,128z};
    \fill[white] svg{M86.3,186.2H70.9V79.1h15.4v48.4V186.2z}
                 svg{M108.9,79.1h41.6c39.6,0,57,28.3,57,53.6c0,27.5-21.5,53.6-56.8,53.6h-41.8V79.1z M124.3,172.4h24.5c34.9,0,42.9-26.5,42.9-39.7c0-21.5-13.7-39.7-43.7-39.7h-23.7V172.4z}
                 svg{M88.7,56.8c0,5.5-4.5,10.1-10.1,10.1c-5.6,0-10.1-4.6-10.1-10.1c0-5.6,4.5-10.1,10.1-10.1C84.2,46.7,88.7,51.3,88.7,56.8z};
  }
}

\renewcommand{\orcidID}[1]{%
  \resizebox{8px}{8px}{
      \href{https://orcid.org/#1}{\tikz[yscale=-1,transform shape]{\pic{orcidlogo}}}}%
}
\begin{document}
\title{Validation-Driven Development\thanks{The research presented in this paper has been conducted within the IVOIRE project, which is funded by ``Deutsche Forschungsgemeinschaft'' (DFG) and the Austrian Science Fund (FWF) grant \# I 4744-N and has been partly financed by the LIT Secure and Correct Systems Lab sponsored by the province of Upper Austria.}}
%
%
\author{
Sebastian Stock\orcidID{0000-0002-2231-8656} \and
Atif Mashkoor\orcidID{0000-0003-1210-5953} \and
Alexander Egyed\orcidID{0000-0003-3128-5427}
}

\authorrunning{Stock et al.}
%
\institute{
Johannes Kepler University\\
Altenbergerstr. 69, A-4040 Linz, Austria\\
\email{\{firstname.lastname\}@jku.at }
}

\maketitle              
\begin{abstract}
Formal methods play a fundamental role in asserting the correctness of requirements specifications. However, historically, formal method experts have primarily focused on verifying those specifications. Although equally important, validation of requirements specifications often takes the back seat. This paper introduces a validation-driven development (VDD) process that prioritizes validating requirements in formal development. The VDD process is built upon problem frames - a requirements analysis approach - and validation obligations (VOs) - the concept of breaking down the overall validation of a specification and linking it to refinement steps. The effectiveness of the VDD process is demonstrated through a case study in the aviation industry. 
\end{abstract}
\keywords{Validation-driven development, validation obligations, formal methods, Event-B}

\section{Introduction}
\label{sec:introduction}
Formal methods play a crucial role when developing critical systems, allowing a correct specification of the system behavior. This specification can be checked for consistency via verification that often takes preeminence in formal development. Consequently, techniques like model checking, theorem proving, and associated toolsets such as SPIN~\cite{holzmann97a} or Isabelle~\cite{paulson94a} are widely used in industry. On the other hand, the compliance of the specification with desired system behavior can be ensured via validation. Validation is supported by techniques like animation and simulation and associated toolsets like AsmetaA~\cite{asmetaa} or JeB~\cite{mashkoor17a}. Contrary to verification, using validation techniques and toolsets is less common, especially in state-based formal methods~\cite{mashkoor18a}. Even if used, they are considered a secondary activity towards the end of the development cycle. 

A typical formal requirements specification process starts with a set of (natural language) requirements. Once specified, requirements undergo a stringent verification process for consistency checking. Then, the validation process follows. The whole development process is iterative. Verification is often given preeminence over validation because it does not make sense to validate something inconsistent. However, prioritizing verification over validation may lead to crucial issues, such as keeping the end users out of the loop. While specifiers create, verify, and validate specifications, the end users only give inputs at the specification process's beginning or end. Consequently, late feedback means more changes, efforts, and costs. 

Many techniques have been proposed to overcome this problem. For example, Baumeister~\cite{baumeister04a} suggested using test-driven development (TDD) for writing formal specifications. The author proposes generating run-time assertions from the specification to check for compliance between specification, code, and tests. Later, Bonfanti et al.~\cite{bonfanti18a} proposed using behavior-driven development (BDD) for a similar cause. The advantage of BDD over TDD is that it supports early collaboration among stakeholders, such as specifiers, developers, quality assurance experts, and end users, by giving low-level tests a high-level meaning. However, this reliance on tests comes with a price, and while testing is a valid means of validation, it is not necessarily exhaustive enough to cover all validation challenges. BDD restricts itself to a scenario language translated to some test in the target language (e.g., natural language to LTL formula), which may not be extensive enough to validate all properties of interest. This translation is often limited depending on the expressiveness of the target language. Furthermore, BDD is usually applied to formal specifications late in the process.

This paper proposes a validation-driven development (VDD) process for writing formal specifications that puts validation at the center of formal development. The VDD process focuses on creating validatable specifications, allowing end users to subjugate the formal specification process. Furthermore, VDD suggests a highly expressive and systematic structuring, elicitation, documentation, tracing, and maintenance process for formal requirements specifications appealing to all stakeholders.

The VDD process is built upon two well-known concepts: problem frames~\cite{jackson2001problem} and validation obligations (VOs)~\cite{mashkoor2021validation}. Problem frames help analyze requirements in a structured and collaborative manner. On the other hand, VOs help check the compliance of a specification concerning the stakeholders' requirements and support incremental specification writing and evolution. Analogous to proof obligations, VOs break down the overall validation of a specification and associate it with the specification's refinement steps.

The rest of the paper is structured as follows: \Cref{sec:background} provides the necessary background to understand the content of this paper by introducing Event-B and VOs. However, note that the findings of this paper are language-independent. \Cref{sec:vdd} introduces and exemplifies the VDD process. \Cref{sec:case_study} demonstrates the application of the VDD process through a case study from the aviation domain. \Cref{sec:releated_work} compares the VDD process to other similar approaches. Finally, \Cref{sec:conclusion} concludes the paper with some proposed future work.

\section{Background}
\label{sec:background}
\subsection{Event-B}
\label{subsec:eventB}
The formal language Event-B~\cite{abrial2010modeling} is based on first-order predicate logic and set theory and helps with specification writing, verification, and proving using the platform Rodin~\cite{abrial10a}. The behavior of a specification is defined using \texttt{machines} that contain a set of \texttt{variables}, which are described in the \texttt{invariant} section. \texttt{Events} are considered state transitions, with a guard marked with the \texttt{when} clause that must be true before enabling the event. \texttt{Context} defines the static part of a specification. The Event-B language supports both vertical and horizontal refinement styles. While vertical refinement is about concretizing the abstract data structure, horizontal refinement is about introducing additional features to the specification. 

\subsection{Validation obligations}
\label{subsec:vo}
\begin{figure}
    \centering
    \includegraphics[scale=0.6]{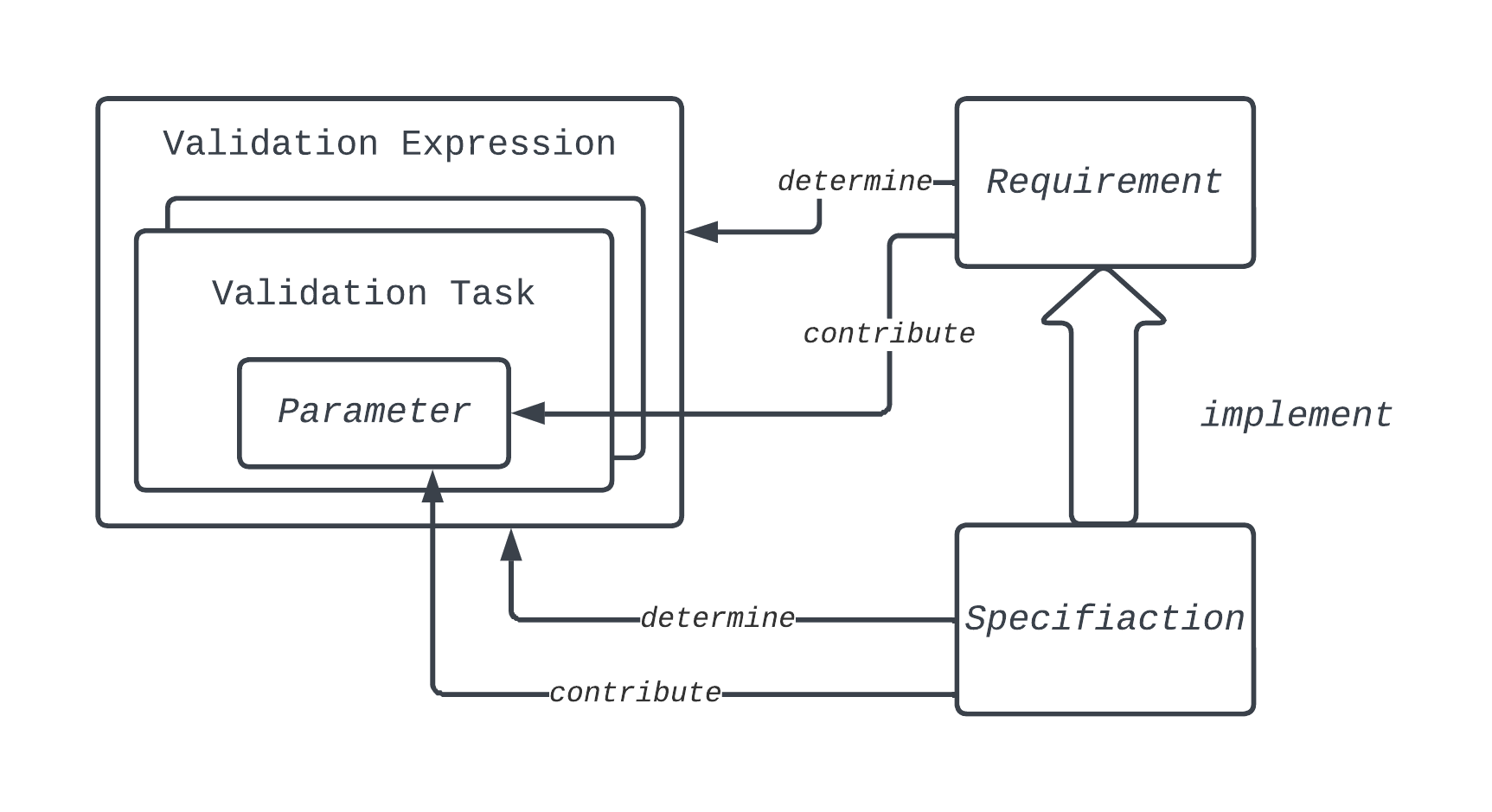}
    \caption{Internal view of a VO}
    \label{fig:vo_workflow}
\end{figure}

Validation obligations (VOs) are logical formulas associated with the correctness claims of given validation properties. Each VO represents a requirement showing evidence of its existence in the specification. \Cref{fig:vo_workflow} shows the internal components of a VO and their interplay. A validation expression (VE) is run against the specification and can consist of one or more validation tasks (VT) connected by the logical operators $\vee, \wedge$, and $;$. The semicolon operator represents a validation expression where the components before and after the semicolon share the same state space of the specification. Thus, this operator allows for complex validation expressions where steps depend logically on each other. Typical validation techniques are animation, simulation, testing, or model checking. VTs have parameters determined by the requirements and structures of the specification. VOs help us with traceability, documentation, and maintenance throughout the specification as they act as tokens documenting how a requirement is realized in a specification. Further, they indicate when a requirement is no longer satisfied. 

Let us consider the following requirement in a lift example where the operator can choose between multiple floors from 0 to 2. \texttt{REQ0:}\textit{The floor level will eventually equal 2}. This requirement is implemented in specification \texttt{M0}. Suppose we choose LTL model checking as a validation technique. In that case, we can encode the requirement into the following VO, where the parameter is an LTL formula:
\begin{align}
\label{eq:vo1}
    \mathtt{REQ0}/\mathtt{M0}:\mathtt{LTL1}:=\mathtt{FG}(\{ x=1 \})
\end{align}

\section{Validation-driven development}
\label{sec:vdd}
VDD proposes a systematic process for requirements elicitation, documentation, tracing, and maintenance during formal developments. In the following, we discuss the workflow of the VDD process, the role of VOs in specification writing, and the structuring of the specification through problem frames.

\subsection{Workflow}
\label{subsec:workflow}
\begin{figure}
    \centering
    \includegraphics[scale=0.5]{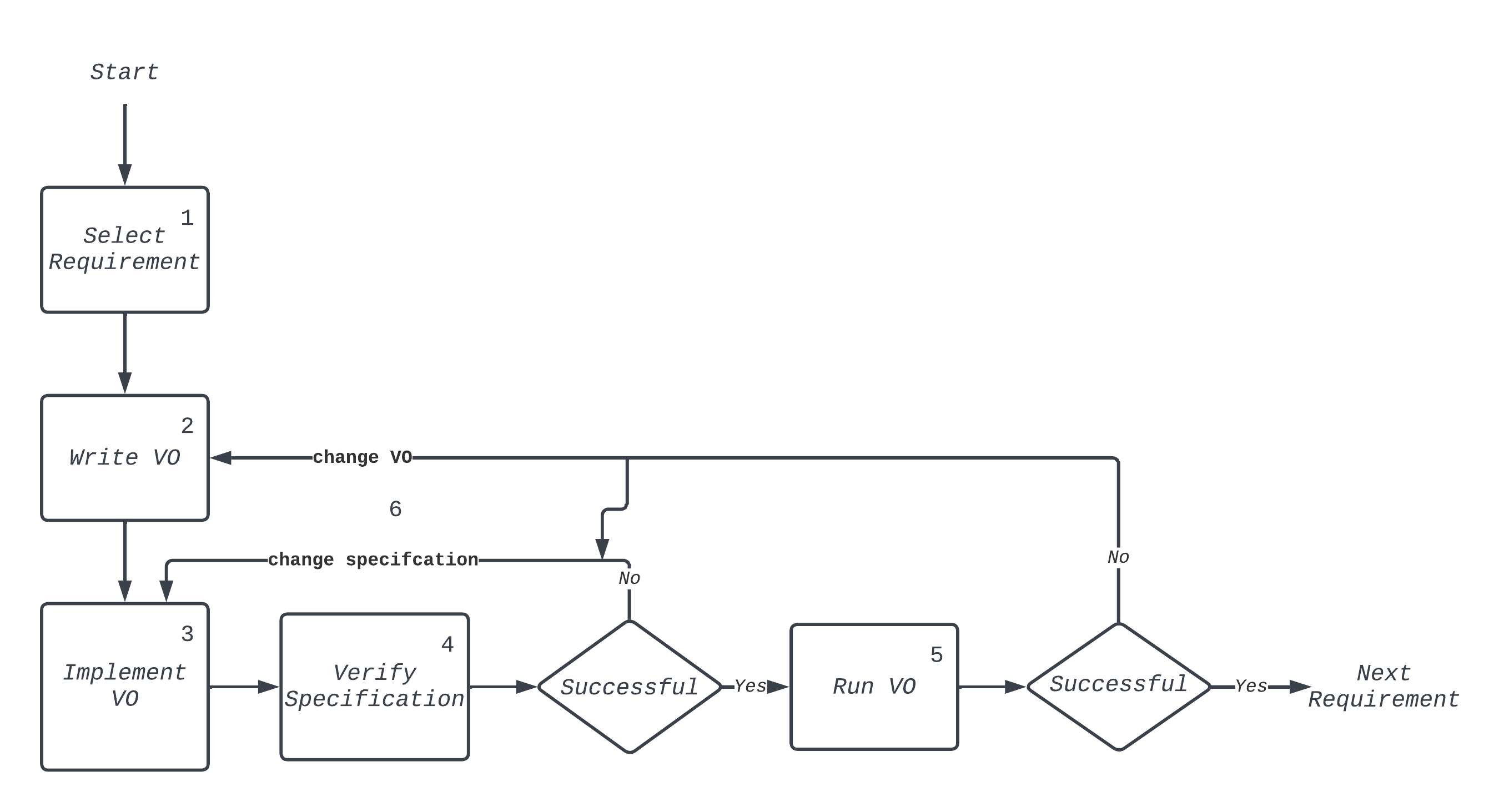}
    \caption{VDD workflow}
    \label{fig:vdd_flow}
\end{figure}

\Cref{fig:vdd_flow} shows the workflow of the VDD process, which is as follows: 

\begin{enumerate}
    \item Select a requirement.
    \item Write a VO that, if successful, would give evidence for correctly implementing the requirement.
    \item Implement the VO in the specification.
    \item Verify the specification, e.g., check for internal consistency.
    \item Run the VO, e.g., execute the associated validation task.
\end{enumerate}
After satisfying a VO, the specifier can introduce a refactoring session to improve the existing specification. Overall, the approach is iterative for all requirements, and if we introduce additional VOs and change the specification, leading to other VOs failing, we have a hint of inconsistency. The VDD process also helps to keep things simple, i.e., if we need to introduce more than a handful of variables, state transitions, and invariants during the VO implementation, we most likely want too much at once. The VO makes this apparent. Checking multiple properties of one requirement hints that the requirement may be divided into sub-requirements. 

\paragraph{Example.}
Let us specify \texttt{REQ0} from the previous lift example (step 1). For this, we first create the VO as shown in~\Cref{eq:vo1} (step 2). Now we need to implement the VO (step 3). We approach this as minimalist as possible. In the LTL formula, we need a variable \texttt{floor} which is some form of a number. Consequently, we start with \texttt{x} equaling 1. Then, we check the specification for internal consistency (step 4). If this is successful, we employ the LTL model checking to evaluate the VO (step 5). 

\subsection{Specification structuring and refinement}
\label{subsec:refinement}
We now focus on problem structuring and refinement planning, two challenging tasks in formal developments~\cite{HALLERSTEDE20142}. The first challenge is to recognize what aspects of the problem are related to which other elements, i.e., eliciting the structure of the specification. The second challenge is to derive a valid refinement structure from this, which supports the verification and validation process. However, both challenges require experience to master them.

\begin{figure}[t]
    \centering
    \begin{subfigure}[b]{0.48\textwidth}
         \centering
    \includegraphics[scale=0.5]{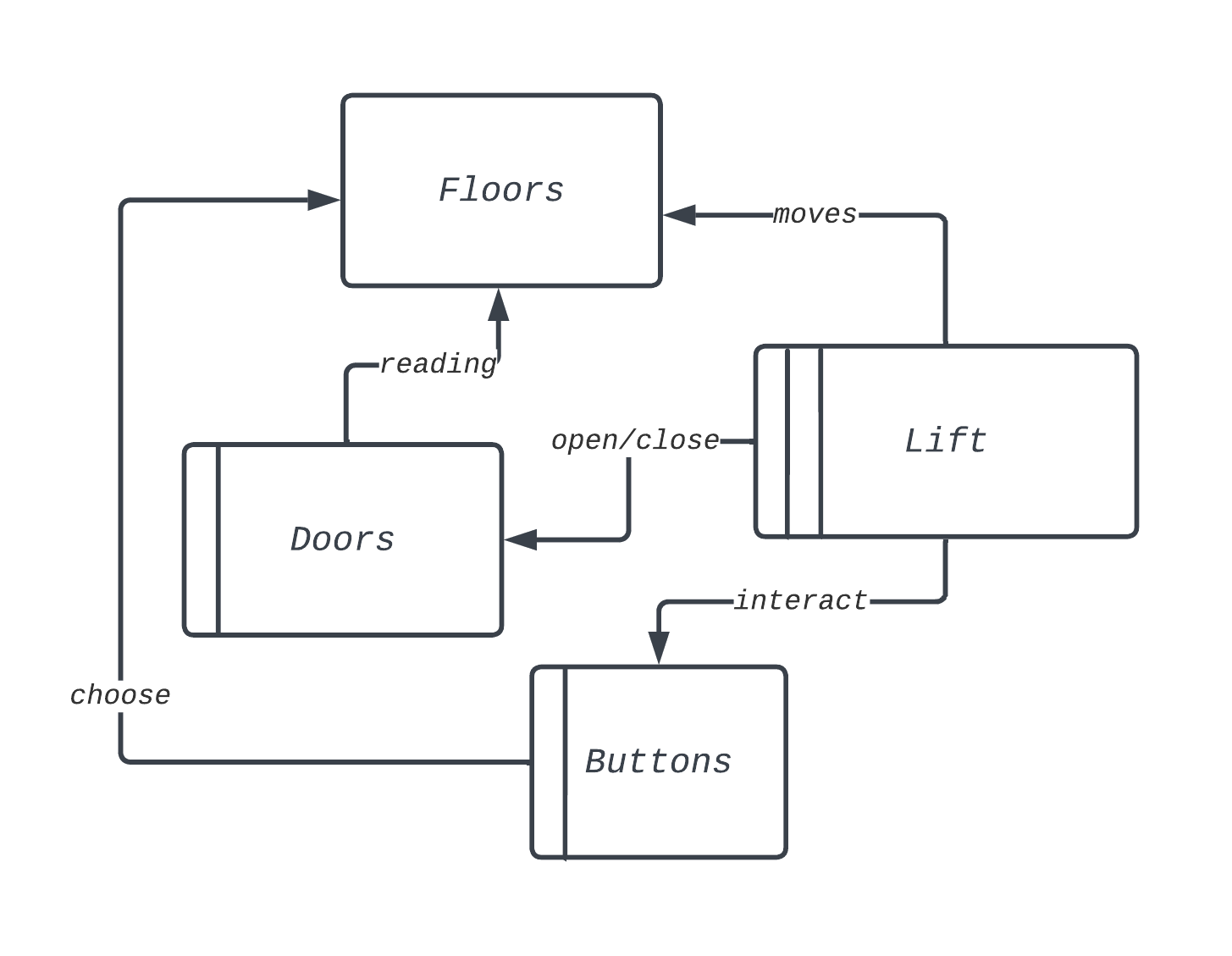}
    \caption{The main problem}
    \label{fig:main_problem_frame}
    \end{subfigure}
    ~
    \begin{subfigure}[b]{0.48\textwidth}
              \centering
    \includegraphics[scale=0.5]{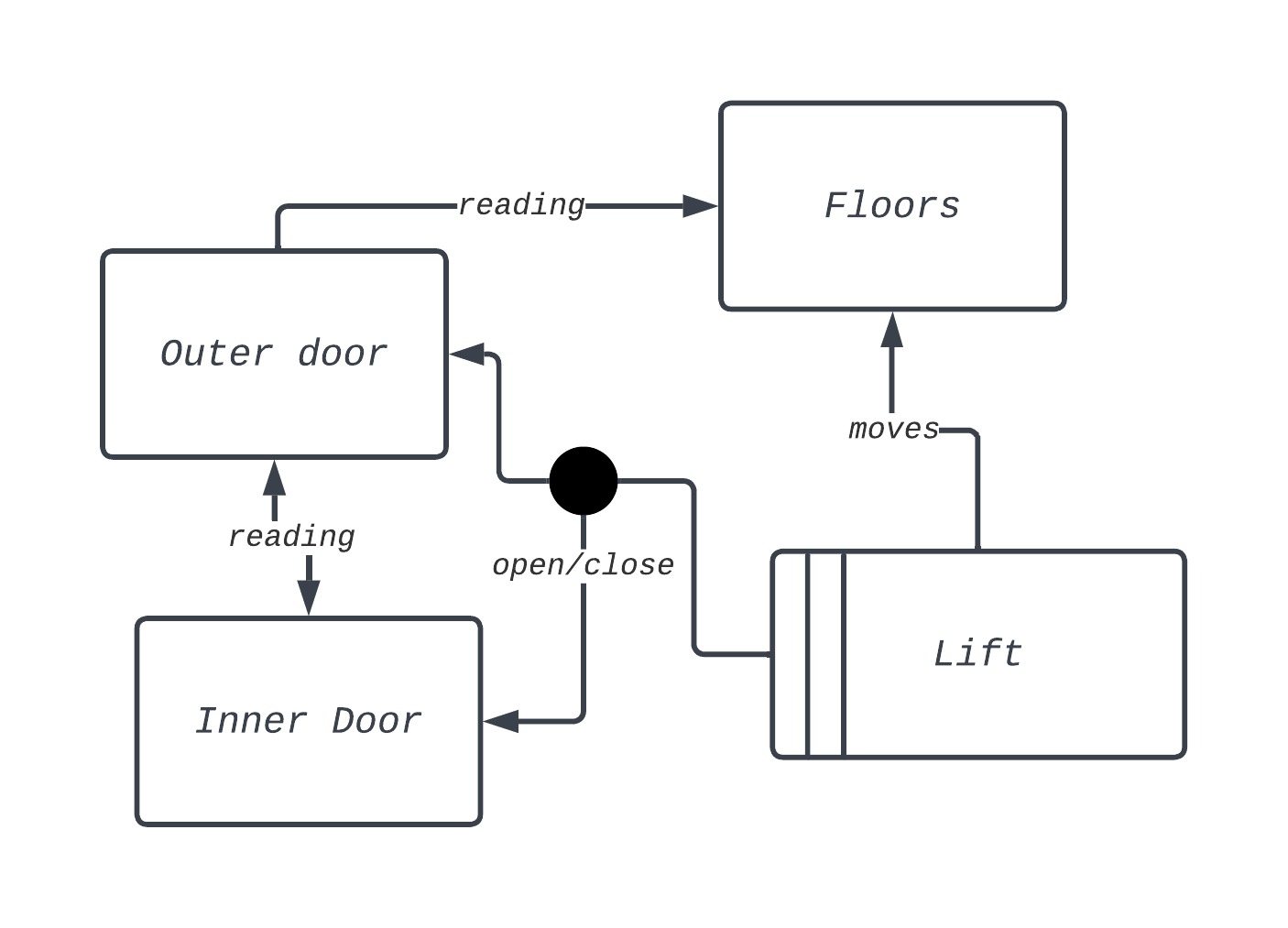}
    \caption{The door sub-problem}
    \label{fig:sub_problem_frame}
    \end{subfigure}
    \caption{Problem frame of the lift problem and the sub-problem concerning \texttt{Doors}}
    \label{fig:problem_frame}
\end{figure}

\paragraph{Framing the problem.}
We adapt the problem frame methodology~\cite{jackson2001problem} to structure specifications. \Cref{fig:problem_frame} shows the problem frame of the lift example. The rectangles represent the concerning domains. Each domain represents an aspect of the physical world that is observable to the stakeholders. Lines between the domains are interfaces and explain how the domains interact with each other. The rectangle with the doubled vertical stripe is the \textit{machine domain}, the specification we want to write. Rectangles with one stripe are \textit{designed domains} that represent the information we are free to express as we desire. This notation is for complex domains where the design is up to the specifier but where the details do not concern the global problem. Finally, rectangles with no stripe are \textit{given domains}. Given domains are those we need to consider but cannot alter their appearance. They are usually very abstract for our specification purposes and require less attention. Our addition to the problem frames is the arrows on the interfaces indicating an information flow. Either they are uni- or bi-directional.

\paragraph{Example.}
In our running example in \Cref{fig:main_problem_frame}, we want to specify a lift with three areas of concern. The \texttt{Floors} we want to navigate to are a given domain we cannot change. The lift \texttt{Doors} need to be detailed and marked as a designed domain. Finally, the \texttt{Buttons} is also a designed domain, as we have yet to get further instructions on how the buttons should look. Going into further detail, in \Cref{fig:sub_problem_frame}, we can see a sub-problem only concerning the lift's \texttt{Doors}. This sub-problem was separated as it would bloat \Cref{fig:problem_frame} with information only specific to one domain. We can see that we replaced the \texttt{Doors} domain with two more specific domains related to each other. The \texttt{Outer Doors} are the doors on each floor. \texttt{Inner Door} is the lift's door and must read the outer door status to synchronize accordingly. Furthermore, three domains share the \texttt{open/close} interface. The arrows at the interfaces show us the dependencies of their interaction, mainly the lift specification. In reality, domains are chosen and marked according to given and extracted information. This can lead to eliciting new requirements to fill gaps between the desired specification and reality. Moreover, we may involve non-technical stakeholders in the process due to a visual structure.

\paragraph{Structuring specification.}
We use the following guidelines to structure the specification:
\begin{enumerate}
    \item Domains sharing an interface will need to interact eventually. Therefore, they should refine each other horizontally (e.g., \texttt{Doors} is dependent on \texttt{Floors} in our problem frame and should refine it).
    \item The first domain to be implemented is the one with the most connecting incoming interfaces. We then implement the domain with the second-highest incoming interfaces and proceed iteratively (e.g., specifying \texttt{Floors} before \texttt{Doors} as \texttt{Floors} has the most incoming interfaces).
    \item Whenever we omit details in the main problem frame and create a sub-problem frame, we are confronted with a choice: 
    \begin{enumerate}
        \item We can introduce the details immediately, substituting the domain with the domains introduced in the sub-problem (e.g., \texttt{Doors} is immediately specified as \texttt{Outer Doors} and \texttt{Inner Door}).
        \item We can introduce the details later in a vertical refinement and keep the abstract domain around (e.g., \texttt{Doors} is refined to \texttt{Outer Doors} and \texttt{Inner Door}).
    \end{enumerate} 
    \item Whenever multiple domains share an interface without being connected otherwise, they may be related in a vertical refinement relationship.
    \item Domains not directly connected to the machine domain are of secondary concern.
\end{enumerate}
Structuring the specification further and fostering understanding for stakeholders involved, we can annotate domains with corresponding requirements extracted from the requirements document. This helps later with the elicitation of VOs. We distinguish between two types of VOs: VOs focusing on the domain and VOs focusing on the interplay of two domains. Separating both concerns helps estimate the validation effort, as VOs focusing on the domain will likely still be valid if we change unrelated domains.

\paragraph{Example continued.}
Applying these guidelines, we can derive a specification structure. For example, to specify the lift, we would start with the floors, as they are referenced most (Guideline 1). Next, we would specify the \texttt{Doors}. Here we are confronted with a choice. We can keep the \texttt{Doors} abstract for now and move on to the \texttt{Buttons} (Guideline 3b) or detail the \texttt{Doors} before moving on (Guideline 3a). The decision for either is dependent on the requirements we want feedback on. If we keep the \texttt{Doors} abstract (Guideline 3a), we can introduce the \texttt{Buttons} and gather early feedback on the whole system and the interaction between domains. On the other hand, if we choose to introduce the details of the \texttt{Doors} (Guideline 3b), we encounter a special case of two domains sharing an interface and being connected independently. The dependency structure is that \texttt{Outer Doors} and \texttt{Inner Door} complement each other as the bidirectional interface indicates. However, as a sub-problem, they refine the \texttt{Doors} domain. Consequently, both domains are introduced at the same time. Therefore the problem frame helped us to evaluate the impact of possible specification structures.

\paragraph{Validation and refinement.}
When introducing VOs early and then applying changes to the specification due to refinement or refactoring, we must tackle the (re)validation question. We can use the problem frame to indicate where revalidation might become necessary. For example, in horizontal refinement relationships, if we have an incoming interface, i.e., we consume information from another domain and change the producing domain, we must revalidate every VO consuming from this producing domain. Analogous is true for having producing domain. Adding to the insights proposed by Stock et al.~\cite{stock2023a} if a VO only concerns a single domain and is not dependent on others, outside changes do not invalidate it. For vertical refinement, rechecking VOs depends on the specification language. If the specification language has a strict notion of refinement, such as Event-B, where we can show the preservation of safety and liveness properties, our VOs will stay intact. For specification languages featuring a liberal notion of refinement, such as ASMs, we might recheck VOs. In some cases, the VO can be transferred, preserving its insights. For example, the works of Arcaini et al.~\cite{arcaini2019} and Stock et al.~\cite{stock22} tackle the problem of information transfer, and the insights can be applied to VOs.

\section{Case study}
\label{sec:case_study}
\subsection{System description}
\begin{figure}
    \centering
    \includegraphics[scale=1.5]{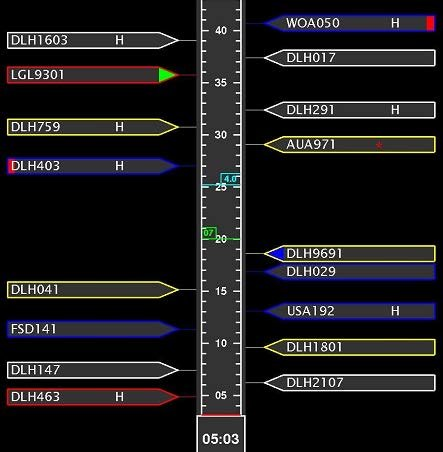}
    \caption{Screenshot of the AMAN system~\cite{martinie2012using}}
    \label{fig:aman}
\end{figure}

We exemplify the VDD process on the Arrival Manager (AMAN) case study~\cite{palanque23a}. The AMAN system focuses on developing a human-machine interface for managing aircraft arriving at an airport. The particularity lies in continuously scheduling new aircraft to land at the airport while users can interact with the schedule on a screen in three different. The first interaction to consider is dragging the aircraft to another landing slot via the mouse. The second is blocking landing slots and disallowing the computer from scheduling aircraft in this slot. The third is to put the aircraft on \textit{hold}, meaning that the countdown till landing is not reduced for these planes. Furthermore, the user can zoom in and out on the landing schedule, thus reducing or increasing the presented slots and aircraft, respectively. \Cref{fig:aman} shows the working of the AMAN system. In the middle, one can see the remaining time till landing, and the boxes on the left and right are planes. Colors indicate different statuses, for example, \textit{hold}.

\subsection{Problem structuring}
\label{subsec:specification_structure}
This subsection demonstrates how the requirements of the AMAN system can be specified using the VDD process. We use the problem frames approach introduced in \Cref{subsec:refinement} to understand and define the problem. For brevity, only a portion of the case study is shown here. For the complete specification, please consider the work of Geleßus et al.~\cite{gelessus23a}. 

\begin{figure}[t]
    \centering
    \begin{subfigure}[b]{0.48\textwidth}
     \centering
    \includegraphics[scale=0.5]{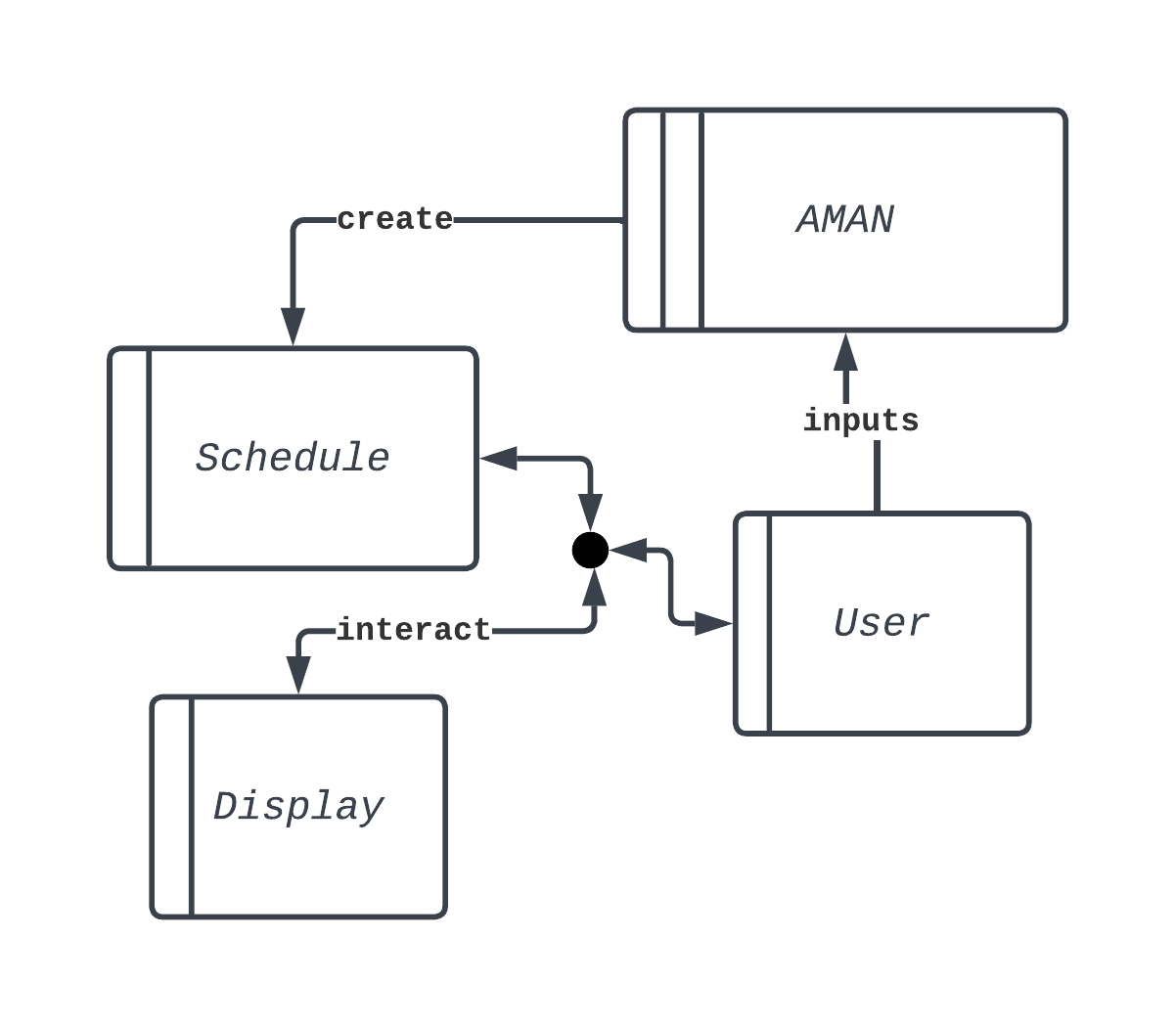}
    \caption{Problem frame of the \texttt{AMAN} }
    \label{fig:global_problem_frame_aman}
    \end{subfigure}
    ~
    \begin{subfigure}[b]{0.48\textwidth}
    \centering
    \includegraphics[scale=0.5]{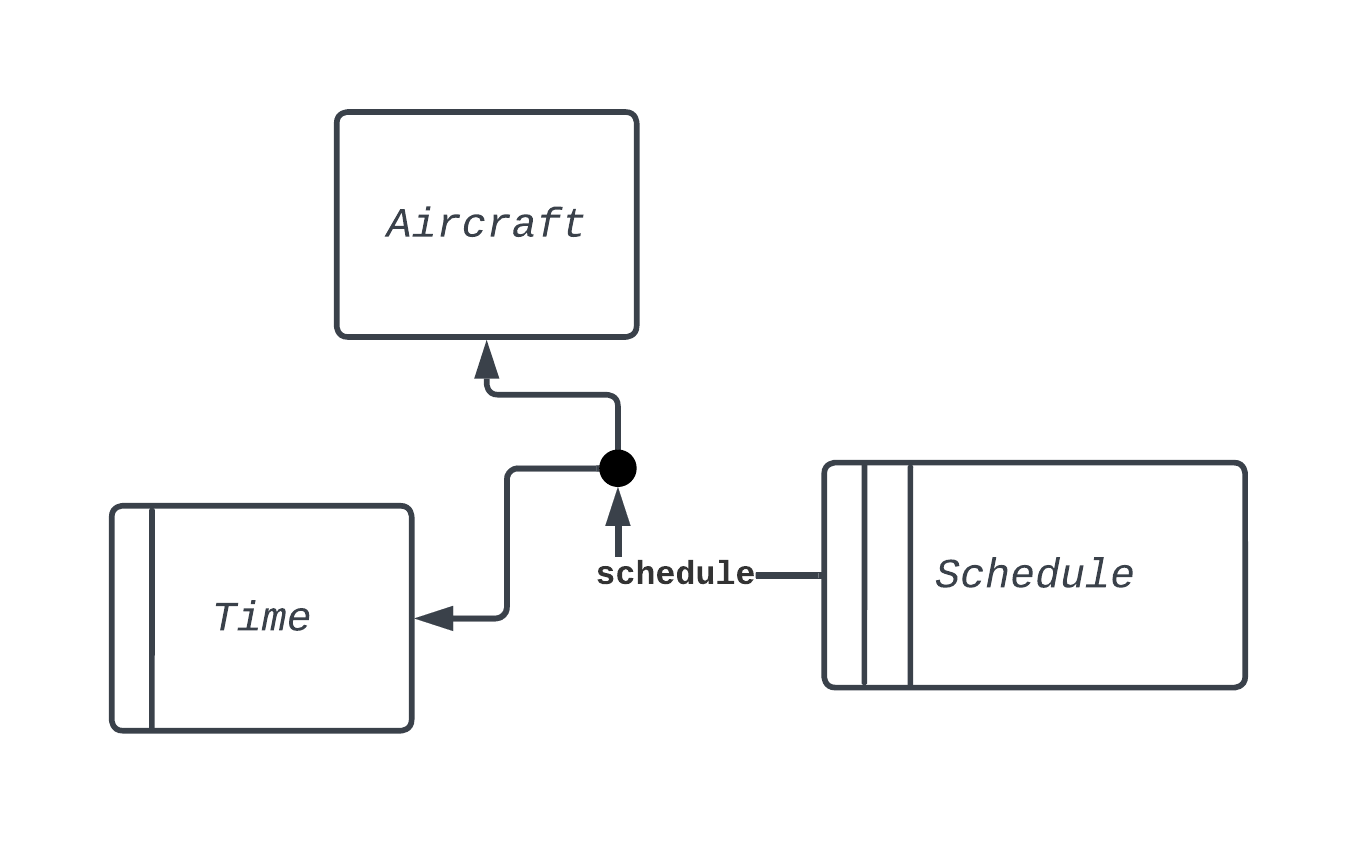}
    \caption{Sub-problem concerning the \texttt{Schedule}}
    \label{fig:sub_problem_frame_aman}
    \end{subfigure}
    ~
    \begin{subfigure}[b]{0.48\textwidth}
    \centering
    \includegraphics[scale=0.5]{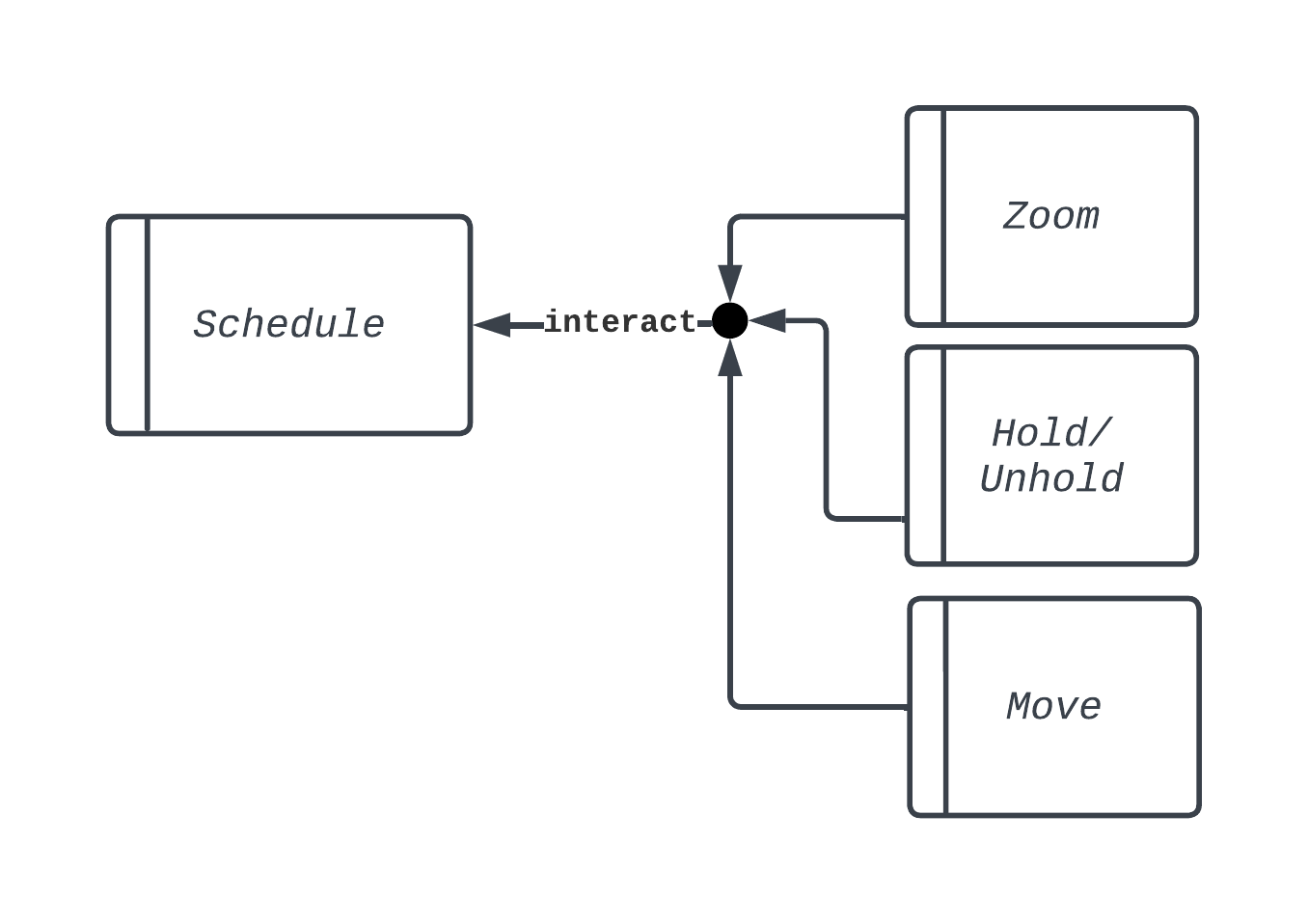}
    \caption{Sub-problem concerning the user interaction}
    \label{fig:sub_problem_frame_aman_user}
    \end{subfigure}
    \caption{Problem frame of the AMAN and a sub-problem frame concerning the scheduling}
    \label{fig:problem_frame_aman}
\end{figure}

\paragraph{Defining domains of interest.} Consider~\Cref{fig:global_problem_frame_aman}, the AMAN we want to specify is marked as the centerpiece by the two extra bars inside; this is the goal of the specification process. Next to the AMAN are designed domains partially mentioned in the system description. Here, we have the designed domain \texttt{User}, which encapsulates the user behavior. For example, the AMAN reacts to the user input. We designated \texttt{User} as a designed domain because we know about some user behavior, but we are unaware of the details and might want to create a sub-problem frame. Then there is the designed domain of \texttt{Schedule}, which encapsulates the process of the AMAN creating a schedule from aircraft and time slots. We marked \texttt{Schedule} as a designed domain as we are not sure of the structure and behavior of the schedule and want to investigate further. Finally, we have the designed domain \texttt{Display} that works as a transmitter as a physical way of transmitting user inputs to the \texttt{Schedule}. However, the lack of an interface with the AMAN suggests its secondary role.

\paragraph{Sub-problem structure.} Diving deeper into the designed domains, we start with the sub-problem shown in \Cref{fig:sub_problem_frame_aman}. Focusing on the \texttt{Schedule} itself, we now consider the \texttt{Schedule}'s two components: \texttt{Time}, which is again a designed domain, and \texttt{Aircraft}, a given domain. We decided here that \texttt{Aircraft} is a given domain as no detail about \texttt{Aircraft} is available. Therefore, we consider it a rather primitive datatype. On the other hand, \texttt{Time} is complex and might require much consideration. Both tie into the \texttt{Schedule} domain, which, according to the proposed guidelines, indicates a refinement. Additionally, both have the same amount of incoming interfaces. Therefore, we can start specifying with any of them. 

The second sub-problem in \Cref{fig:sub_problem_frame_aman_user} covers the topic of user interaction. Here the domain structure is simple. However, all sub-domains need the \texttt{Schedule}, and additional domains share the interaction interface, which indicates some interference in the domains. Otherwise, the domains remain very loosely connected. What could be a consideration is that we define abstract \texttt{User} interaction that interacts with the \texttt{Schedule} and later refines the \texttt{User} interaction into the three subdomains. This, again, depends on how we define the scheduling.

\paragraph{Final specification structure.}
We can use the proposed guidelines discussed in~\Cref{subsec:refinement} to derive a specification structure from these initial problem frames. Considering incoming interfaces, starting with the \texttt{Schedule} seems reasonable. We must decide if we detail the \texttt{Schedule} before implementing \texttt{User} interaction. An argument for this would be that we can validate the most basic function of the AMAN and get feedback on it. Further, we tackle the difficult representation of time early. Afterward, we may implement the \texttt{User} interaction. We subjugate the choice of what to implement first to what needs the most investigation and validation effort, as the individual \texttt{User} interactions only are loosely connected. Finally, we can conclude with the specification of the \texttt{Display} properties. The \texttt{Display} has no direct connection to the primary concern of the AMAN system. Therefore, its specification is a secondary concern.

The final specification structure is as follows:
\begin{enumerate}
    \item Create the \texttt{Schedule} (Guideline 2):
        \begin{enumerate}
            \item Introduce the \texttt{Aircraft} domain (Guideline 3a)
            \item Vertically refine the created specification by introducing \texttt{Time} (Guideline 3a \& 4)
        \end{enumerate}
    \item Horizontally refine the specification by introducing \texttt{User} interactions (Guideline 1 \& 3a) and consequently \texttt{Zoom}, \texttt{Hold/Unhold}, and \texttt{Move} in any order (Guideline 1)
    \item Horizontally refine the specification by introducing \texttt{Display} (Guideline 5)
\end{enumerate}

\subsection{Specification and validation}
We start the specification process with the \texttt{Schedule} sub-problem. In the following, we refer to requirements directly derived from the specification as a direct quote: \texttt{REQX} with \texttt{X} being a number. According to the tactics presented in \Cref{subsec:workflow}, we start by selecting a requirement, creating a VO, and then specifying the requirement. For example, let's assume we select the requirement of \texttt{REQ1:} ``Planes can be added to the flight sequence, e.g., planes arriving in close range of the airport.'' This requirement means: a) we have aircraft, b) we have something to store them, and c) we can manipulate this storage by adding planes. Let's formulate this as a VO:

 \begin{align*}
     \mathtt{REQ1}/\mathtt{M0}:\mathtt{GF}(\mathtt{BA}({\mathtt{scheduledAirplanes} \neq \mathtt{scheduledAirplanes\$0}})) \implies \\ 
     \mathtt{GF}(\mathtt{BA}(\{\exists x.(x \in \mathtt{scheduledAirplanes} \land x \notin \mathtt{scheduledAirplanes\$0})\}))
 \end{align*}

The \texttt{GF} (Globally-Finally) operator indicates that the brackets' expression will eventually be true. The \texttt{BA} is the \texttt{before-after} operator, comparing the current version of a variable with the previous version marked with an \texttt{\$0}, i.e., the difference between \texttt{scheduledAirplanes} in one step and the next step is observed. The LTL formula will ensure that our scheduled aircraft can contain an aircraft not previously in the set of scheduled aircraft. This, however, implies some state transition in our specification, going from an initial state to a state with one more aircraft that was not previously contained.

\begin{figure}[t]
         \centering
    \includegraphics[scale=0.4]{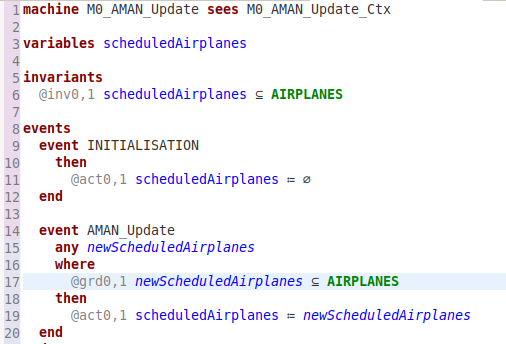}
    \caption{The schedule sub-problem with only aircraft}
    \label{fig:m0_imp}
    \end{figure}

\Cref{fig:m0_imp} is an Event-B specification that attempts to satisfy the VO. We have a variable representing our \texttt{Schedule}, an \texttt{AIRPLANE} datatype, and an event creating a new schedule, eventually satisfying the VO. We could now generate more VOs to ensure soundness implementation regarding the amount of added planes. For now, we are satisfied and proceed. 

Taking a look back at \Cref{fig:sub_problem_frame_aman}, we need to implement the \texttt{Time} domain to cover the \texttt{Schedule} domain fully. The corresponding requirement we want to satisfy by introducing the time is \texttt{REQ5:} ``The space between two aircraft is always $\geq$ 3, with 3 being the time in minutes.'' Following is the corresponding VO. 

 \begin{align*}
     \mathtt{REQ5}/\mathtt{M1}: &\forall \mathtt{a1},\mathtt{a2} \cdot \mathtt{a1} \in \mathtt{dom}(\mathtt{landing\_sequence}) \land \\
  			&\mathtt{a2} \in \mathtt{dom}(\mathtt{landing\_sequence}) \land \mathtt{a1} \neq \mathtt{a2} \implies \\
           &(\mathtt{DIST}(\mathtt{landing\_sequence}(\mathtt{a1}) \mapsto \mathtt{landing\_sequence}(\mathtt{a2})) \\
           &\geq \mathtt{AIRCRAFT\_SEPARATION\_MIN})
 \end{align*}
 
\begin{figure}[t]
    \centering
    \includegraphics[scale=0.4]{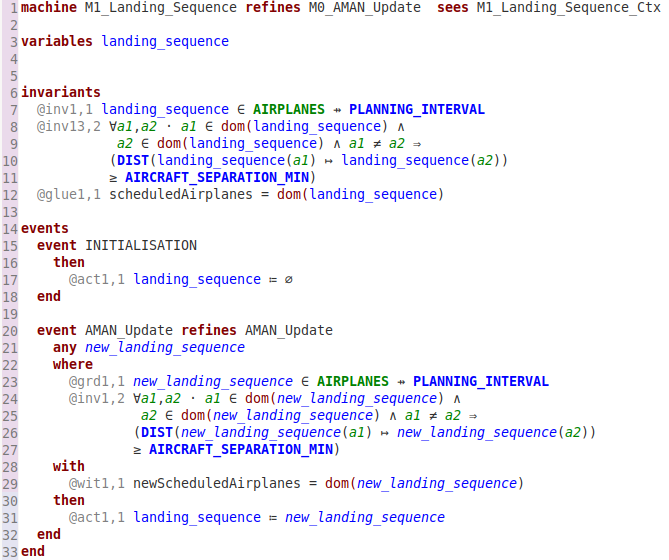}
    \caption{The schedule sub-problem with added time}
    \label{fig:m1_imp}
\end{figure}

For this VO, we assumed that we upgraded our \texttt{scheduledAirplanes} from \Cref{fig:m0_imp} to \texttt{landing\_sequence} as shown in \Cref{fig:m1_imp}, which is a mapping from aircraft to time slots. Consequently, we demand that every aircraft contained in this mapping has a distance (\texttt{DIST}) to every other aircraft of \texttt{AIRCRAFT\_SEPARATION\_MIN}, which in our case is 3. Consequently, we must upgrade our \texttt{scheduledAirplanes} and take care of the proof. 

\Cref{fig:m1_imp} shows the corresponding specification. We introduced the mentioned \texttt{landing\_sequence} and further introduced \texttt{inv13,2} to establish proof. Furthermore, we refined our \texttt{event} to use the upgraded data structure. After discharging the proof, we establish that our requirement is truly represented in the specification.

As previously established, both domains \texttt{Aircrafts} and \texttt{Time} have a connection, and therefore when creating \texttt{M1}, we need to show that \texttt{REQ1} is still preserved in the specification. As Rodin only supports a safety-preserving notion of refinement, re-establishing the VO must happen by re-executing the LTL formula.

After completing the scheduling sub-problem, we move on to the \texttt{User} interaction part. Our VOs concerning the \texttt{Schedule} will not be revalidated when validating the interaction. We only consume the \texttt{Schedule}'s behavior as laid out at the end of \Cref{subsec:specification_structure}. 

\section{Related work}
\label{sec:releated_work}
Several approaches have been proposed for the validation of requirements specifications. While some focus on the whole specification process, others focus only on certain aspects. We briefly introduce and compare some of them with our proposed process.

\subsection{BDD usage in formal requirement specification}
\label{subsec:bdd}
BDD~\cite{carlosBDD} is a well-established technique in the area of software development. It is appealing due to its easy-to-follow procedure and its effectiveness in establishing that requirements are part of the code. First, a scenario is created and run (with intermediate steps) against the code. If this is successful, the next scenario is tackled. If it fails, either code or scenario has to be fixed. One strength is the imposed iterative nature, which comes naturally by adding more satisfied scenarios. Furthermore, tracing and maintaining requirements is massively simplified as every requirement has one scenario mapped to a group of tests. Naturally, attempts have been made to use BDD in formal developments.

There are many significant adaptations of the BDD approach for the formal specification community. For example, Snook et al.~\cite{snook2018behaviour} proposed an Event-B targeting version of Cucumber~\cite{wynne2017cucumber} to describe scenarios in the Gherkin\footnote{\url{https://cucumber.io/docs/gherkin/}} language which is translated into a trace and executed against a specification. The scenario language FRETISH~\cite{giannakopoulou2020generation} goes in a similar direction as it can be used to express requirements which are then converted to an LTL formula with the help of the FRET~\cite{giannakopoulou2020formal} tool. This approach orients itself heavily on what Gehrkin does for programming examples. It provides a basic language to write scenarios, which can be (automatically) linked to LTL formulas. 

While these approaches can be applied successfully, they suffer from two drawbacks. First, they consider validation after writing specifications, thus losing out on the advantages of validation-centered specifications. Doing validation last will compromise completeness due to time constraints or the complexity of the specification. Second, scenario language used in BDD causes problems of expressiveness and, therefore, suffers from a lack of completeness. Second, while these approaches work well, they only provide one solution to a validation problem. We must rely on the correct translation from the scenario language to the validation technique. Furthermore, there is no way to choose between different validation techniques to translate the scenario. This means a method like FRETISH can only react to a scenario by producing an LTL formula. However, model checking may not always be a good solution, e.g., in infinite state spaces. 

VDD addresses both concerns while keeping the compact and easy-to-follow style of BDD. First, it puts validation at the center of the formal development process. Second, it offers a liberal syntax allowing for expressing and consequently validating different properties of interest with many techniques and tools.

Arcaini et al.~\cite{arcaini2019} showed how BDD-like scenarios targeting ASMs can be transferred between refinement steps of abstract state machines. While the previously mentioned disadvantages to using BDD-like scenarios apply, this work highlights the importance of the transferability of validation results. In the context of VDD, with our approach, we know early when results are transferable or might be due to revalidation, as pointed out at the end of \Cref{subsec:specification_structure}.

\subsection{Bridging the gap between natural language requirements and formal specification}
Several efforts have been made to narrow the gap between natural language requirements and formal specifications, as it can reduce the mental load placed on the specifier, and it helps when attempting to involve non-technical stakeholders. The efforts are bidirectional: creating specifications from natural language requirements and validating natural language requirements in specifications. As discussed in \Cref{subsec:bdd}, BDD for formal specifications caters to the latter concern. 

Regarding creating specifications from natural language requirements, Golra et al.~\cite{golra2018bridging} focus on creating intermediate steps with meta-models for systematically translating requirements to formal specifications. A second work of Sayer et al.~\cite{sayar2019bridging,sayar2020formalization} uses translation patterns. However, as both approaches introduce intermediate layers of abstraction, they also introduce additional error sources where the translation could be wrong. Furthermore, they may suffer from the same problems discussed in \Cref{subsec:bdd}, where the intermediate language might not be powerful enough to translate the constructs. 

VDD does not introduce intermediate layers but changes the standard order from specification first to validation. Therefore no new error source was introduced. Furthermore, the mental load is reduced as the problem is tackled in smaller portions. Finally, with VOs, non-technical stakeholders can get a feeling for the progress the specification made and point to requirements that still need work.

\subsection{Requirements tracing}
Another field of interest is systematically tracing the implementation status of requirements. Exculpatory for these efforts are, for example, the works \cite{HALLERSTEDE20142,jastram2010approach}, where a sophisticated set-theoretic representation for requirements is proposed, which is supposed to help with the tracing of requirements. Compared to our approach, the authors heavily focus on the properties of Event-B and proofing with proof obligations. Validation is a gap filler for everything that cannot be proven. While our work also contributes to traceability, it takes a more lightweight approach inspired by software development strategies and thus is more intuitive. Furthermore, the focus is on validating and creating validatable specifications, not fitting a validation solution to an existing specification.

\section{Conclusion and future work}
\label{sec:conclusion}
This paper presents the validation-driven development process for writing formal specifications. It offers an iterative approach to formal specifications, strongly focusing on their validation. The aim is to provide a systematic process to structure, elicit, document, trace, and maintain formal requirements specifications. To this end, we employ an adapted version of problem frames complemented by validation obligations.

In the future, we want to provide tool support that helps automate the VDD process by keeping track of VOs, the specification structure, and changes. Especially the steps of VOs elicitation and creation could be fully automated.

\bibliographystyle{splncs04.bst}
\bibliography{bib.bib}
\end{document}